# Application of a miniaturized photoacoustic cell for high-sensitivity laser detection of ammonia in gas media.

A.V. Gorelik, V. S. Starovoitov

*B.I.Stepanov Institute of Physics, NASB, Nezavisimosti ave. 68, 220072 Minsk, Belarus*
*e-mail: a.gorelik@dragon.bas-net.by, vladstar@dragon.bas-net.by*

**Abstract.** We present a photoacoustic resonant cell, the internal volume of which is ~ 0.5 cm$^3$. The cavity shape of the cell is so adapted in order to minimize the background signal arising due to absorption of laser beam in the cell windows. An experimental setup, measurement procedure and design of the cell are described. Results of detection of ammonia in nitrogen flow with the help of a CO$_2$-laser are represented. The minimal detectable absorption for the cell is estimated in the experiment to be ~ $5.1 \cdot 10^{-8}$ cm$^{-1}$ W Hz$^{-1/2}$.

Laser-based photoacoustic (PA) spectroscopy is one of the most sensitive techniques applied to local non-contact analysis of trace amounts of various chemical compounds in gas medium [1,2,3]. The most appropriate light sources for this technique are lasers operating in the infrared wavelength region. They are powerful carbon dioxide (the oscillating wavelength region 9 - 11 μm) and carbon monoxide (2.8 - 4.1 or 5.1 - 8.0 μm) CW lasers [4,5], CW parametric oscillators (3 - 5 μm) [6], quantum cascade lasers (4 – 5 μm) [7,8] and miniature DFB laser diodes (0.7 - 2 μm) [9]. For instance, CO$_2$-lasers can be applied successfully in order to detect selectively compounds incorporating such chemical bonds as C-O, C-N, C=S, P=O, C-F, S=O, Si-O-, Si-H, P-OC, P-OH, S=O, N-H, P-H, As-H and others [10]. The use of rare isotopic species in the laser active medium allows both to widen the oscillation wavelength range (that is, to increase the number of substances to be detected) and to reduce essentially (more than 100 times) a deleterious effect of the background absorption in atmospheric air on the CO$_2$-laser-based measurements [11,12,13,14].

The PA technique is distinguished by a high sensitivity at the (sub) ppb (parts per billion, $1:10^9$) level with the minimal detectable absorption of $10^{-10}$ cm$^{-1}$, a short time resolution (down to a few seconds) and the capability to detect a large number of chemical compounds [2,3]. The mid-infrared-laser-based PA technique is generally recognized to be a reasonable approach to the inexpensive *in situ* multi-component analysis of atmospheric air in environmental pollution monitoring [2,15,16], exhaust gas monitoring [2,17,18], industrial process control [19,20] and leak detection [21]. The technique finds an expanding application in biology-related areas (plant physiology [3], crop and fruit storage [22], entomology [23], medical diagnostics by means of exhalation analysis [24,25]) for sensitive detection of CO$_2$, C$_2$H$_4$, CH$_4$, NO$_2$, N$_2$O, H$_2$S, O$_3$, NH$_3$, C$_2$H$_6$, C$_2$H$_6$O, C$_5$H$_{12}$ and others gases being involved in the metabolic reactions for living systems.

The miniaturized spectroscopic hardware of trace-gas analysis is a promising field of application for the technique. The principle of the technique is based on measuring the amplitude and phase for the acoustic pressure oscillation (the so-called photoacoustic response) arising due to absorption of a modulated laser beam by molecules of gas inside a photoacoustic cell. The photoacoustic detection is realized with an enhanced sensitivity if the modulation frequency coincides with an acoustic resonance of the cell. Therefore, the modulation frequency should be correlated with the cell-resonator sizes: the frequency increases with reducing the sizes. Implementation of laser-beam modulation at ultrasonic frequencies ranged from tens kilohertz up to the submegahertz scale allows to reduce the cell sizes down to a few millimeters. Combining the high sensitivity inherent in the traditional PA-approaches with an ability to probe the gas inside such a small-sized PA cell gives a possibility to analyze chemical compounds to be emitted by individual small-sized objects with an extremely low emission rate. A crude estimation shows that the application of the approach can provide detecting the gas leak emitted by an object with the rate down to ~ $10^{-14}$ cm$^3$/s for $C_2H_4$ or ~ $10^{-10}$ cm$^3$/s for $CO_2$. For comparison, in the aerobic reaction an individual cell of living organism emits $CO_2$ with a rate from $10^{-10}$ to $10^{-9}$ cm$^3$/s [26]. In the photosynthesis reaction an individual cell of plant can absorb $CO_2$ at a rate higher than $10^{-8}$ cm$^3$/s [27]. The detection sensitivity will be increased $10^3$ – $10^6$ times compared to the traditional PA approaches and more than 100 times in relation to the commercial mass-spectrometer-based leak-detection systems. In contrast to the systems, the ultrasonic PA leak detector needs no expensive vacuum chambers and can be applied to *in situ* localization of leak for a large number of substances to be emitted in atmospheric air. According to the theoretical estimations (see, for instance, [28, 29, 30]), the amplitude of photoacoustic response can be enhanced with reducing the cell sizes.

The quartz-enhanced photoacoustic spectroscopy (QEPAS) is a reasonable approach to miniaturize the spectroscopic measurements with the help of the increased-frequency modulation of laser beam [31]. Instead of a gas-filled resonant acoustic cavity, the sound energy is accumulated in a high-Q quartz crystal frequency standard. Usually the standard is a quartz tuning fork with an acoustic resonant frequency of 32 kHz in air. To the moment, a great success is achieved in QEPAS-based gas detection with the help of compact cells (the internal cell volume reaches down to a few cubic centimeters) and different near- or mid-infrared laser systems (including diode and quantum cascade lasers, optical parametric oscillators) [32,33,34]. The experiments testify that the high-sensitivity photoacoustic detection can be realized at ultrasonic modulation frequencies if the rates of intra- and intermolecular collisional vibration-vibration (VV) or vibration-translation (VT) energy redistribution for the detected species are high compared to the modulation frequency.

The goal of the report is to demonstrate the application potential for another possible approach to the miniaturization of photoacoustic hardware. In contrast to the QEPAS technique, we support a traditional approach to photoacoustic detection: through a small hole in the cell shell, a condenser microphone registers the photoacoustic response to the laser beam modulated at the frequency of an acoustic resonance of the cells. We accept that the miniaturized photoacoustic cell must not be worse in the performance in comparison to non-miniaturized one. The cell performance is specified usually in terms of the signal-to-noise ratio. In order to obtain a high magnitude for the ratio the cell design must satisfy some requirements. First of all, the design has to provide detection of the highest possible useful signal (that is, a photoacoustic response from the gas inside the cell) at a selected acoustic resonance. The cell must be reliably isolated from the external acoustic noise and parasite electric pickups. And, finally, the background effect of the cell windows (or, sometimes, of the internal cell surface) on the measurements should be reduced to a minimum.

This background effect is associated usually with absorption of laser beam in the cell windows resulting to local heating of the neighboring gas inside the cell. The measurement inaccuracy arisen from this effect is proportional both to the laser power instability and to the power absorbed by the windows. If a non-miniaturized photoacoustic cell is used the effect can be critical for high-power laser applications (for instance, at an intracavity arrangement of the cell to a carbon dioxide or carbon monoxide laser operated with the intracavity power of some tens Watts). The background effect is not very significant for non-miniaturized low-power laser systems (such as laser diodes or quantum cascade lasers operated at the power up to a few tens of milliWatts). But, our experience testifies that the size reduction for the photoacoustic cell results in a dramatic increase in the background signal and this signal can be observed for a low-power laser beam. For instance, the window background signal for a small-sized photoacoustic cell is observed in our experiments at a laser beam power down to 10 mW. We assume that an efficient way to minimize the window background effect and to enhance the cell performance can be implemented with the help of a properly adapted shape for the cavity of photoacoustic cell.

In the report we present experiments on laser detection of gases. In the experiments, a photoacoustic resonant cell (the internal volume of which is less than 1 $cm^3$) of a simplified cavity shape is represented. We demonstrate how such a shape can be applied to minimize background signals arising due to absorption of laser beam in the cell windows. An experimental setup, measurement procedure and design of the cell are described in brief. Results of detection of ammonia (a fast relaxing chemical compound) with the help of a $CO_2$-laser are represented.

The experimental setup is shown in Figure 1 to consists of four main functional parts: a gas-generator system (includes an air/nitrogen generator 1, humidity generator 2, thermo-diffusion generator 3 and flow controller 4), laser system (carbon dioxide laser 12, unit for precision wavelength-tuning 11 driven by computer-controlled device 14 and system of active laser stabilization 15), computerized system of experiment control and signal processing (personal computer 16, acousto-optical modulator 10, digital oscilloscope 13, power meter 9, photodetector 8) and a photoacoustic cell 7.

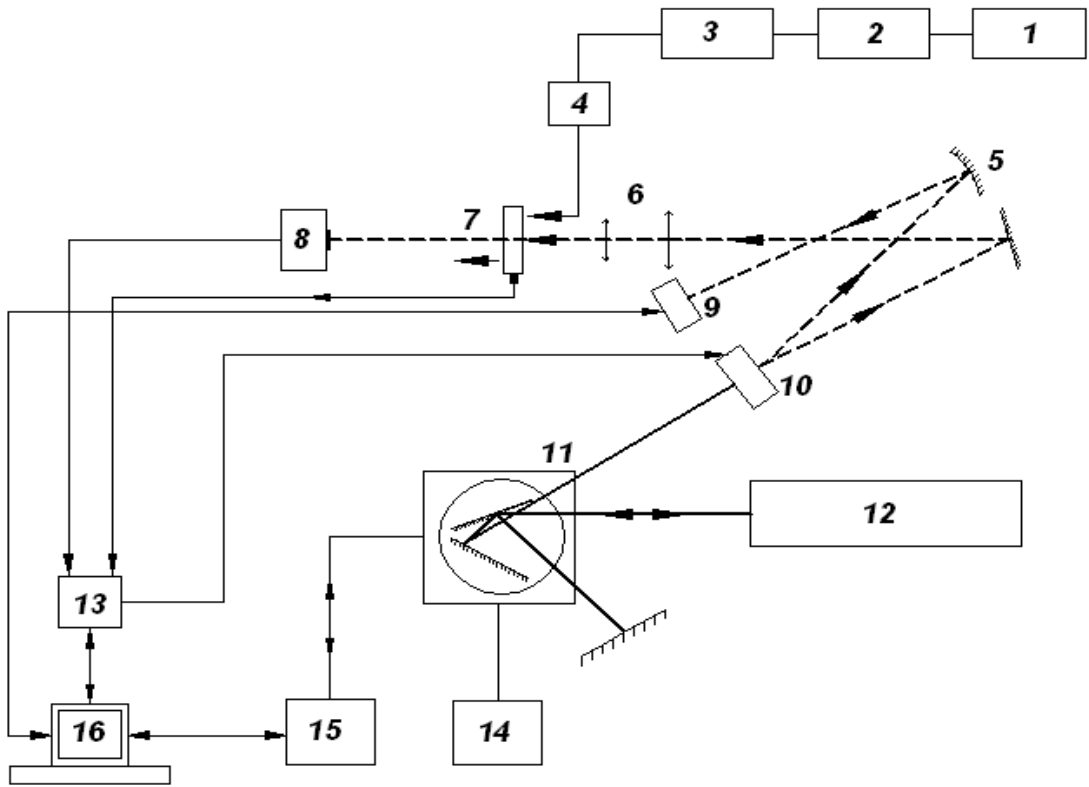

Figure 1. Functional scheme of experimental setup

The gas-generator system provides the calibration gas flow for the measurements. The heart of the system is a generator ANG250A which extracts ultra-high-purity nitrogen (purity > 99.9995 %) or clean dry air (the air purity corresponds to a dew point of -70 C) from ambient laboratory air by 'Pressure Swing Adsorption' (PSA) method. The humidity generator VX-G maintains a required humidity level for the nitrogen/air flow and delivers water vapor to the flow at a dew point from 0.2 to 85 C. The thermo-diffusion generator GDP102-2 is a constant temperature system designed to generate the small amount of a chemical compound (for instance, ammonia at ppm/ppb concentrations) in the gas flow, using permeation devices as trace-gas sources. The gas-flow rate is maintained with the help of a flow controller RRG-33 (the full scale and accuracy are 10 and 0.2 cm$^3$/min, the time response < 0.1 s).

Advantages of our laser system are the compactness (the laser cavity length is 96 cm) and opportunity to apply a computerized wavelength tuning. The system is capable also to operate at auto-collimating or non-auto-collimating oscillation regimes. The auto-collimating regime (the so-called

Littrow scheme) gives the highest possible power for the laser lines. The non-auto-collimating one (the Littman scheme) provides both the reliable selection of a laser line and a stable single-mode operation of the laser. The principal part of the system is a stabilized continuous wave carbon dioxide laser (sealed-off discharge tube LG-205) tunable over the oscillation wavelengths in the range from 9 to11 µm. The wavelength tuning is performed by a diffraction grating (150 grooves/mm, the blazing angle of 31 angular degree, the reflection to the zero and first orders of diffraction are of 3 and 94 %, correspondingly) mounted in a piezoelectric ceramic holder. The holder is attached rigidly to a corner reflector. The reflector is turned with the help of a mechanic rotation stage 8MR180 driven by a step motor Phytron. The motor operation is controlled by a computer interface 8CMC1-USBh. The accuracy of the corner turning is accepted to be limited by the angular displacement (4.5 angular seconds) for one step. The stabilization of laser oscillation is performed at an active regime by tuning the laser-cavity length with the help of a commercial frequency-stabilization unit LG-74. The stabilization principle is based on the optic-galvanic effect. The aperture of the output laser beam is ~ 2 mm. For individual oscillation lines, the output power ranges from 200 mW (the non-auto-collimating regime) to ~ 1 W (the auto-collimating regime).

The probing laser beam is modulated by a Germanium acousto-optic modulator (the efficient aperture is 3 mm, the efficiency of beam deflection is 80 %). The deflected beam is directed to a power meter Ophir 3A by spherical mirror 5. In the following (for instance, when we estimate the cell performance), we accept the power of the deflected beam as an actual power of the probing beam. The deflection angle for the beam depends on the oscillation wavelength $\lambda$. Therefore, the radius of the mirror is so selected in order to minimize the possible transverse beam displacements relative to the power meter. The non-deflected beam (it does not depend on $\lambda$) is directed trough a collimating system 6 (Germanium lenses with anti-reflection coating) and the photoacoustic cell to a photodetector. The transmission factor for the lenses is higher than 95% at $\lambda \sim 10.6$ µm. The factor of beam reduction for the lens system is 4. The aperture of the laser beam near the photoacoustic cell is estimated to be 0.5 mm. The deflection of the laser beam by the acousto-optic modulator is switched with the help of a TTL signal from a signal generator built in a digital oscilloscope Handyscope HS3. The switching frequency is accepted to be the frequency of beam modulation. The digital oscilloscope is used in order to register, store and transmit the obtained electric signals (the 'optical' signal from photodetector and 'acoustic' signal from the microphone built in the photoacoustic cell) to the computer. The amplitude and phase for the detected photoacoustic signal are found from the ratio between the Fourier transforms obtained at a modulation frequency for the optical and acoustic signals.

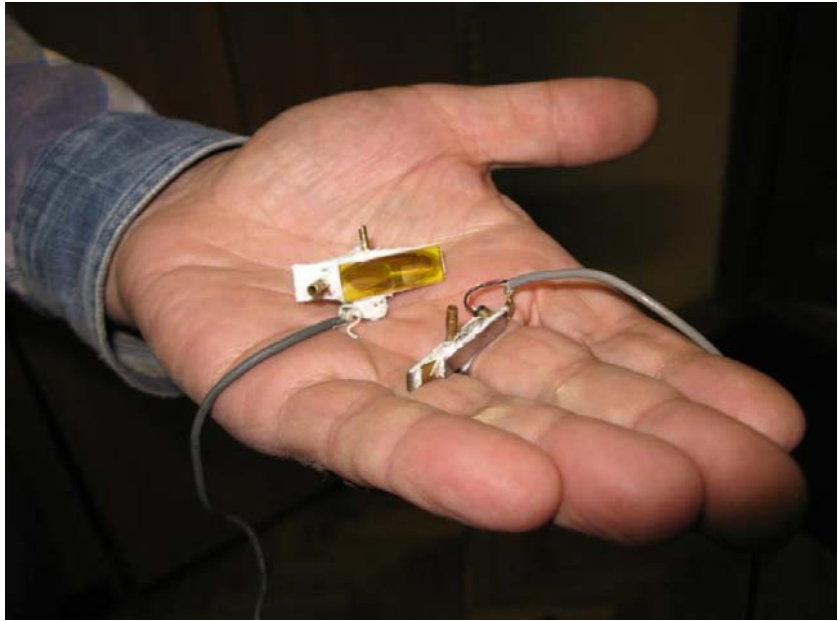

Figure 2. A set of miniaturized photoacoustic cells with inclined-ring geometry. The left cell is a ZnSe-windowed (internal volume V ~ 500 mm$^3$) cell represented in the report. The right one is a cell (V~80mm$^3$) with Ge windows.

We test a photoacoustic cell of simple design. Geometrically, the cell is similar to a ring inclined to the laser beam at the Brewster angle (see photo in Figure 2). The parameters, which specify the shape of the cell cavity (they are the diameter of clear aperture and thickness of the ring; the location of microphone and inlet/outlet ports), are fitted in order to enhance the cell performance at an individual acoustic resonance. The resonance corresponds to the acoustic mode $\nu_2$ of the cell and is specified by the second low-lying eigen-frequency. The parameters are fitted with the help of a numerical simulation based on the finite difference and finite element methods. In the simulation the spatial amplitude distribution for a stable acoustic standing wave (the wave corresponds to the mode $\nu_2$) inside the cell is determined. From this distribution we find both the useful signal to be generated by gas inside the cell and background signal caused by absorption of the laser beam in the cell windows. The simulated signals are analyzed as functions of the geometry parameters. The parameters are fitted in such a way as to increase the useful signal and, simultaneously, to minimize the background signal for the mode $\nu_2$. The cell windows are made of polycrystalline ZnSe. The internal cell volume is ~ 0.5 cm$^3$ and the diameter of clear aperture is 8 mm. The length of optical path inside the cell is no longer than 1 cm. A miniature condenser Knowles FG microphone is used for detection of the acoustic signals. Our numerical simulation predicts that in the frequency range from 0 to 20 kHz there are three acoustic modes of the cell: $\nu_1$, $\nu_2$ and $\nu_3$. According to the simulation, the eigen-frequency is located near 14.2 kHz.

The miniaturized cell is applied in a demonstration experiment. In the experiment we detect ammonia in nitrogen flow on individual oscillation lines of carbon dioxide laser. The rate of flow to be blown through the cell is maintained automatically with the help of the flow controller in such a way to provide a perfect gas renewal inside the cell for the time ~ 1 sec. The pressure in the cell is accepted to be close a typical pressure for the environment (740 torr). In order to provide a reliable selection of laser lines the measurements are performed when the laser operates at the non-auto-collimating regime. The laser *'on-line'* wavelength to be used for ammonia detection corresponds to the vibration-rotation line 9R(30) of band $00^01$-$[10^00,02^00]_{II}$ of $^{12}C^{16}O_2$ molecule. The absorption $\alpha_{on}$ on this line for ammonia is 52.6 cm$^{-1}$atm$^{-1}$. Such a strong absorption implies that the photoacoustic signal detected on the line can be accepted reliably as a useful signal from the gas to be analyzed. The effect of the parasite signals in the detection is estimated from the data obtained by measurements on the line 9R(24) (*'off-line'*) of the same band. The absorption on this line for ammonia is negligible small (0.006 cm$^{-1}$ atm$^{-1}$) in order to have any influence on the detected signal. Indeed, our experiment testifies that the signal detected on the line does not depend on the content of ammonia in the flow. We recognize this signal therefore as a manifestation of the window background or the noise.

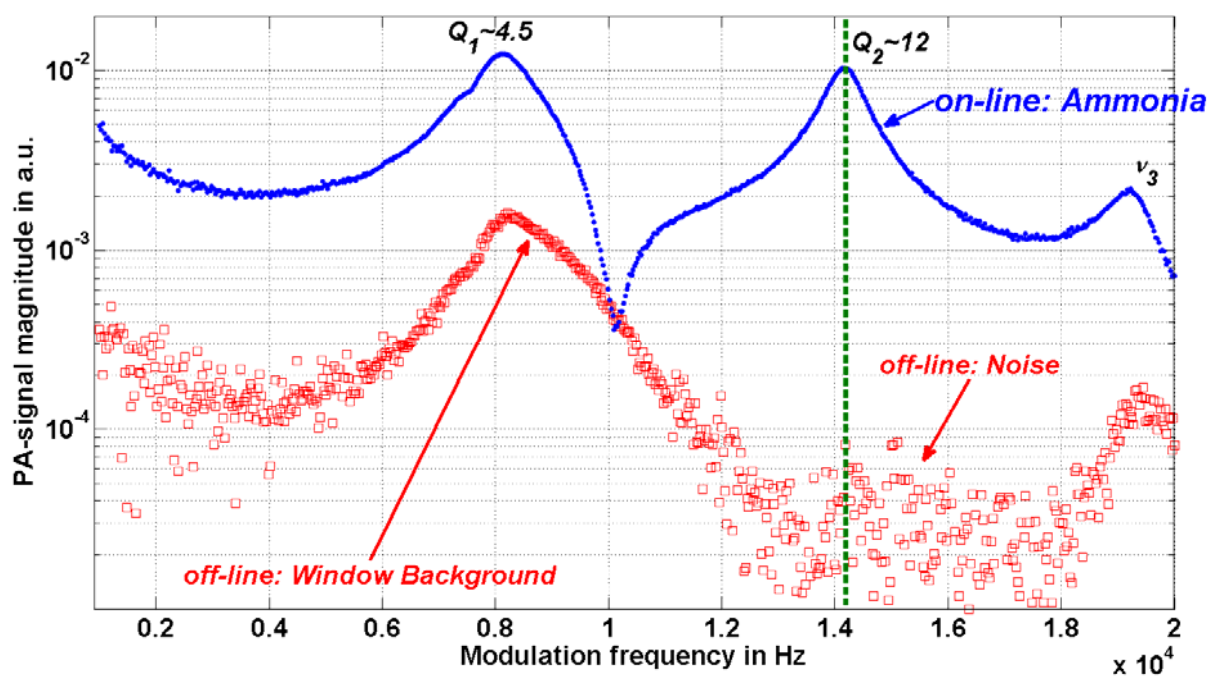

Figure 3. The dependence of the *on-line* (blue full circles) and *off-line* (red open squares) photoacoustic signals on the modulation frequency ω. The measurements are made at the ammonia concentration of 13.6 ppm in nitrogen flow. The power of the *on-line* or *off-line* beams is 87 or 171 mW correspondingly. The averaging time for every frequency point of the dependence is 0.13 sec.

To test the cell capabilities we analyze the *on-* and *off-line* dependences of the signal amplitude and phase on the modulation frequency ω ranging from 1 to 20 kHz. In the absence of ammonia in the flow, the dependences are identical. If ammonia is admixed to the flow the *'on-'* and *'off-line'* signals taken separately exhibit the different behavior on the frequency scale. The behavior of the *on-line* dependence is considerably transformed and the signal amplitude is increased. The *off-line* dependence remains invariable. Figure 3 shows the frequency dependences of signal amplitude. The *'on-line'* signal demonstrates clearly the resonances $\nu_1$, $\nu_2$ and $\nu_3$ predicted by the numerical simulation. The detected photoacoustic signals shows the corresponding resonant peaks at frequencies $\omega_1 = 8.133$ kHz (the Q-factor for the resonance $Q_1 = 4.6$), $\omega_2 = 14.2$ kHz ($Q_2 = 12.3$) and $\omega_3 = 19.2$ kHz. Contrary to the *'on-line'* dependence, the *'off-line'* signal manifests only two resonant peaks. They are resonances at frequencies $\omega_1$ and $\omega_3$. There is no resonance peak in the vicinity of $\omega_2$. Moreover, near $\omega_2$ the *'off-line'* signal is strongly noised and the signal amplitude demonstrates a minimum comparable with the standard deviation (the root-mean-square error) obtained for the detected acoustic signals in the absence of laser beam. The analysis of the *'off-line'* signal measured at the modulation frequency $\omega = \omega_2$ shows that the signal phase is distributed in a quasi-uniform random manner over the range from 0 to $2\pi$. It implies that the magnitude for the window background at $\omega = \omega_2$ is negligible small in comparison to the noise. Such an effect testifies to the capability to minimize efficiently the window background for the acoustic mode $\nu_2$ by optimizing properly the cell shape.

Obviously, the performance of the cell is maximal when the modulation frequency ω is close to $\omega_2$. Therefore we estimate the cell performance at this frequency. The performance is specified in terms of the minimal absorption $\alpha_{min}$, which can be detectable at the laser power of 1 W and the averaging time of 1 sec provided that the signal-to-noise ratio is equal to 1.

$$\alpha_{min} = \alpha_{on} \, C_{NH3} \, \tau_{avr}^{1/2} \, P_{on} \, N_{off}/S_{on}$$

Here the quantities $C_{NH3}$ and $\tau_{avr}$ and denote actual values for ammonia concentration $C_{NH3}$ (13.6 ppm) and the averaging time (0.13 s). The power $P_{on}$ is found from the *on-line* power for the beam deflected by acousto-optical modulator ($P_{on} = 87$ mW). The parameter $S_{on}$ gives the amplitude for the on-line signal at $\omega = \omega_2$. The quantity $N_{off}$ corresponds to the root-mean-square error for the acoustic signal detected in the off-line measurements. The minimal detectable absorption $\alpha_{min}$ is estimated to be ~ $5.1 \, 10^{-8}$ cm$^{-1}$ W Hz$^{-1/2}$.

Thus, we have presented a simple design for the cell cavity. The shape is specified by a restricted number of parameters, which can be optimized in order, for instance, to adapt the cell geometry to the experiment needs and to enhance the cell performance. Certainly, the shape simplicity makes the parameter optimization and cell manufacturing easy and extends the area of cell application in practice. The cell sizes can be reduced hereafter by a trivial scaling procedure. The obtained results testify to a great potential for the miniaturized photoacoustic cells of the inclined-geometry in creating compact ('pocket-sized') high-sensitivity laser sensors of chemical compounds. Such a millimeter-sized cell powered, for instance, by a minute DFB laser diode with a typical output power of ~ 20 mW will allow to detect trace gases at a sensitivity level, which can be attained by the absorption spectroscopy techniques along an optical path of 4 km. The miniature background-free photoacoustic cell can find also a fascination application in combination with a high-power laser for detection of small gas leaks. Unfortunately, to the moment, the performance of our miniaturized cells is not high in comparison to the relevant parameter for non-miniaturized ones. Nevertheless, we expect to enhance the cell performance in the nearest future.

The authors acknowledge A. Palanetski for the help in performing the experiments. This work was performed in the framework of B-1252 project supported by International Science and Technology Center.